\documentclass[submission,copyright,creativecommons]{eptcs}

\usepackage{breakurl}             

\usepackage[usenames,dvipsnames]{xcolor}

\usepackage{amsmath,amssymb,amsfonts,mathrsfs,stmaryrd}
\usepackage{xspace}
\usepackage{wrapfig}

\usepackage[inline]{enumitem}

\usepackage{tikz}
\usepackage{groove2tikz}
\usetikzlibrary{calc,matrix,positioning,fit,arrows,automata,shapes.misc,shapes.geometric,shapes.arrows,shapes.callouts,shapes.symbols,backgrounds,decorations,decorations.pathmorphing}

\usepackage[font=sl]{caption}

\definecolor{darkred}{rgb}{0.55, 0.0, 0.0}

\usepackage{listings}
\usepackage{float}
\floatstyle{plaintop}
\newfloat{Listing}{tb}{prg}
\newcommand{\gts}{GTS\xspace}
\newcommand{\scoop}{SCOOP\xspace}
\newcommand{\corescoop}{CoreSCOOP\xspace}
\newcommand{\groove}{GROOVE\xspace}

\newcommand{\cpm}{CPM\xspace}
\newcommand{\cpmoo}{CPM+OO\xspace}

\newcommand{\ZZ}{\ensuremath{\mathbb{Z}}\xspace}

\newcommand{\BB}{\ensuremath{\mathbb{B}}\xspace}

\newcommand{\Meth}{\ensuremath{{M}eth}\xspace}

\newcommand{\eifblue}[1]{\lstinline{#1}}

\newcommand{\onlinerepo}{\cite{Referee-Webpage}\xspace}

\newenvironment{inparaenum}{\begin{enumerate*}[label=\emph{(\roman*})]}{\end{enumerate*}}

\newcommand{\blue}[1]{\textcolor{blue}{#1}}
\newcommand{\green}[1]{\textcolor{ForestGreen}{#1}}
\newcommand{\gblue}[1]{\textcolor{blue}{\sf #1}}
\newcommand{\ggreen}[1]{\textcolor{ForestGreen}{\sf #1}}

\providecommand{\event}{GaM 2015} 
\title{Towards  Practical Graph-Based Verification for an Object-Oriented Concurrency Model}
\def\titlerunning{Towards  Practical Graph-Based Verification for an Object-Oriented Concurrency Model}
\author{Alexander Heu{\ss}ner
\institute{University of Bamberg, Germany}
\and
Christopher M. Poskitt \quad Claudio Corrodi\\ Benjamin Morandi
\institute{Department of Computer Science\\
ETH Z\"{u}rich, Switzerland}
}
\def\authorrunning{A. Heu{\ss}ner, C.M. Poskitt, C. Corrodi, and B. Morandi}
\def\copyrightholders{A. Heu{\ss}ner, C.M. Poskitt, C. Corrodi \& B. Morandi}

\begin{document}
\maketitle

\begin{abstract}
  To harness the power of multi-core and distributed platforms, and to make the development of concurrent software more accessible to software engineers, different object-oriented concurrency models such as \scoop have been proposed. Despite the practical importance of analysing \scoop programs, there are currently no general verification approaches that operate directly on program code without additional annotations. One reason for this is the multitude of partially conflicting semantic formalisations for \scoop (either in theory or by-implementation). Here, we propose a simple graph transformation system (\gts) based run-time semantics for \scoop that grasps the most common features of all known semantics of the language. This run-time model is implemented in the state-of-the-art \gts tool \groove, which allows us to simulate, analyse, and verify a subset of \scoop programs with respect to deadlocks and other behavioural properties. Besides proposing the first approach to verify \scoop programs by automatic translation to \gts, we also highlight our experiences of applying \gts (and especially \groove) for specifying semantics in the form of a run-time model, which should be transferable to \gts models for other concurrent languages and libraries.
\end{abstract}

\section{Introduction}

\paragraph{Background}
Multi-core and distributed architectures are becoming increasingly ubiquitous, as the focus for delivering computing performance shifts from CPU clock speeds---now reaching their natural limits---to concurrency. Harnessing this power, however, requires a fundamentally different approach to writing software; developers must program with concurrency, asynchronicity, and parallelism in mind. Traditionally this has been achieved through threads, synchronising via low-level constructs like locks and semaphores. This approach, while still pervasive, is difficult to master and notoriously error prone; deadlocks, data races, and other concurrency faults are all-too-easy to introduce, yet are challenging to detect and debug.
In an effort to alleviate this task for programmers, a number of high-level libraries and languages have been proposed that provide simpler-to-use models of concurrency. Examples include Grand Central Dispatch \cite{GCD-Reference} and \scoop \cite{Nienaltowski07a}, both of which support asynchronous concurrent programming through abstractions that are safer and simpler to grasp than threads. The concurrency mechanisms of \scoop, for example, exclude data races by design. Despite such abstractions, programs may still exhibit rich, complex behaviours that are difficult to fully comprehend through testing alone. There is a pressing need for formal models of these systems to facilitate reasoning, comparisons, and understanding, as well as to bring them within reach of current verification tools and techniques.

\paragraph{Initial Problem}
The intricate features of these libraries and languages---including locking, waiting queues, asynchronous remote calls, and dynamic and automatic thread generation---lead to formal models with verification decision questions (e.g.~deadlock detection and the verification of temporal, behavioural properties) that are undecidable.  Existing approaches to tackle this theoretical challenge fall mainly into two categories: verification algorithms working on restrictions to simple approximations, e.g.\ by extended automata models or Petri nets~\cite{Geeraerts-HR13a,BaldanCK08}, or semi-algorithmic approaches on models that try to cover the original features as faithfully as possible, e.g.~by bounded model checking~\cite{DelzannoT13}.

In the context of \scoop---a high-level object-oriented concurrency model, implemented as an extension to Eiffel---there are currently no analysis or verification approaches that work directly on a program's source code without additional annotations.
Recent first steps into the analysis and prevention of deadlocks in \scoop are either based on checking Coffman's deadlock conditions on an abstract semantic level~\cite{Caltais-Meyer14a}, or require code to be annotated with locking orders~\cite{West-NM10a}.
In earlier work~\cite{Brooke-PJ07a}, \scoop programs were translated by hand to models in the process algebra CSP for e.g.\ deadlock analysis; but these models were too large for the leading CSP tools to cope with, and required a new tool to be custom-built for the purpose (which is no longer maintained today).
No further verification approaches for behavioural properties, e.g.~specified in some temporal logic, exist yet.

In addition, these concurrent libraries and languages often have semantics that are not fully formally specified, or are associated with multiple semantics---whether existing as formal specifications or implicitly, by implementation.
The choice of the ``right'' semantic formalisation, however, is a substantial prerequisite for the analysis and verification of a program's source code.
\scoop, for example, has at least four established, different semantic formalisations~\cite{West-NM15a,Morandi-SNM13a,Brooke-PJ07a,Ostroff-THS09a}. This ``semantic plurality'' is an additional source of complication for verification approaches, such as the one we propose in this paper.

\paragraph{Our Approach}
As a first step, we develop---from the core of the language up---a formal model permitting the simulation and verification of \scoop programs. The rich semantic features of \scoop regarding concurrency, (basic) object-orientation, and especially asynchronicity are grasped with the help of graph transformation systems (\gts) that are parameterised by different underlying semantic variants. We also supply a compiler to automate the task of generating input graphs from \scoop source code. These are then analysed with the help of GROOVE, a state-of-the-art \gts tool, which already includes basic model checking algorithms for \gts.

\paragraph{Contribution \& (Closely) Related Work}
The contribution of the paper is thus manifold: first, we provide a formal \gts-based model that covers \scoop's basic features and can be seen as a new, additional operational semantics for the language.
Second, this \gts-model can also be seen as a new general run-time environment for analysing and verifying object-oriented concurrent programs that share \scoop's main features, including approximations of \scoop.
Third, the given analysis approach serves as a first step towards a general framework for verifying concurrent asynchronous programs by also highlighting modelling best practices, which can be transferred to the analysis and verification of other libraries, e.g.~Grand Central Dispatch, in a similar way. Combining all these aspects, we provide, to our knowledge, the first approach for verifying a subset of \scoop programs on the code level with respect to behavioural specifications---including deadlock freedom.
Only the advanced typing mechanisms and some Eiffel-specific features of \scoop are currently out of reach for our automatic verification approach.

For the broader verification community, this paper demonstrates how a \gts-based semantics and tool can be effectively used to model, simulate, and facilitate verification for a concurrent programming language that abstracts away from threads and has a ``frequently evolving'' run-time. For the graph modelling community, this paper presents our experiences of applying a state-of-the-art \gts tool to a non-trivial and practical modelling and verification problem.

The two closest related works are~\cite{Brooke-PJ07a} and~\cite{Morandi-SNM13a}, which both share our first step of providing a new operational semantics for \scoop. Whereas the former formalises the semantics with the help of a process algebraic model in CSP, the latter defines a semantics based on rewriting logic in Maude. Relying on ``classical'' process algebra, the expression of real asynchronicity between concurrent threads and asynchronous remote method calls are not fully supported by the CSP model---contrary to the model we propose. The comprehensive Maude formalisation is currently seen by the community as the gold standard for \scoop and coined our understanding of \scoop's semantics; our model, in contrast, focuses more on the core asynchronous and concurrent features of \scoop, but can be extended to capture the advanced language features inherited from Eiffel (cf.~later comparison in Section~\ref{sec:fullscoop} for details). Both the CSP and Maude models were used successfully to resolve ambiguities in the original, informal descriptions of \scoop's semantics, but are insufficient for general verification tasks. Directly harvesting, for example, the more expressive and complete Maude implementation for deadlock analysis does not scale on even toy examples like the Dining Philosophers program (presented later).

\paragraph{Plan of the Paper}
After introducing \scoop's main concurrency features (Section~\ref{sec:scoop}), we present a formal model which for the sake of simplicity, ignores ``local'' object-orientation and corresponds to a subset of \scoop that we will call \corescoop (Section~\ref{sec:cpm}). We show how to render \corescoop programs as \gts models (Section~\ref{sec:withgroove}). Afterwards, we describe how we extended our \gts model for \scoop to include full object-orientation, and present a workflow for translating \scoop programs into \groove models (Section~\ref{sec:fullscoop}). The latter then allows us to verify programs written in \scoop with the general algorithms already implemented in \groove (Section~\ref{sec:verif}). We conclude with a comparison to related work on \gts-based verification of concurrent object-oriented systems (other references to related work are stated in the corresponding sections) and provide an outlook on our current research.


\section{\scoop: A Concurrent Asynchronous OO Model}
\label{sec:scoop}
\paragraph{\scoop}
\emph{Simple Concurrent Object-Oriented Programming} (\scoop) \cite{Nienaltowski07a,Morandi-SNM13a} is a concurrent, asynchronous, and object-oriented programming model that---with its intricate semantics---provided the motivation and challenge for the work in this paper. The most thorough implementation of the model is as an extension to Eiffel, but it has also been explored within the context of Java \cite{Torshizi-OPC09a}; we shall focus on the former, and take ``\scoop'' in the following to be a synonym for both the model and this principal implementation.

In \scoop, every object is \emph{handled} by a \emph{processor}, a concurrent thread of control with the exclusive right to call methods on the objects it handles. In this context, object references may point to objects handled by the same processor (\emph{non-separate} objects), or to objects handled by other ones (\emph{separate} objects). Given an object reference \lstinline{x} and a method \lstinline{m} that is a \emph{command} (i.e.~does not return a result), a method call \lstinline{x.m} is executed synchronously if \lstinline{x} is non-separate. If \lstinline{x} is separate, then the handler of \lstinline{x} is sent a request to execute the method asynchronously. This latter case is the main source of concurrency in \scoop programs, which is based essentially on message passing between processors.

The possibility of an object having a different handler is captured in the type system by the keyword \lstinline{separate}. In order to prevent data races, calls to a separate object \lstinline{x} are only allowed if the current object's processor holds a lock on the handler of \lstinline{x}. The programmer does not manage these locking requirements explicitly, but rather expresses them implicitly in the formal argument lists of methods: if the arguments of a method contain separate objects, then the objects' handlers will all be locked (simultaneously, atomically, and automatically---at least conceptually) before the method is executed, and released when it is finished.

\paragraph{Dining Philosophers Example}
A simple example highlighting the intricacies and expressiveness of \scoop is an implementation of the Dining Philosophers problem: a number of philosophers sit at a round table that provides only single forks between adjacent pairs, and these philosophers must concurrently and correctly alternate between eating and thinking. The caveat of course is that a philosopher may only eat if they hold both the fork to their left and the fork to their right, and algorithms must ``pick up'' the forks in such a way that prevents a cyclic deadlock from arising. Consider Listing~\ref{lst:dining_philosophers}, which contains an excerpt from the \lstinline{PHILOSOPHER} class of a well-known \scoop solution (available at~\cite{Referee-Webpage}). Each philosopher and fork object is handled by its own processor. Upon creation, each philosopher is ``launched'' by calling the (argumentless) \lstinline{live} method, causing them to concurrently begin the process of eating and thinking. To eat, a philosopher calls the \lstinline{eat} method, passing the separate object references for the two forks. Eating does not commence until the handlers of these forks are locked by the philosopher's handler; conceptually, this occurs simultaneously, avoiding the possibility of deadlock from e.g.~every philosopher locking their left forks only and then waiting indefinitely on their right ones.

\begin{Listing}[!tb]
\caption{Snippet of the \lstinline{PHILOSOPHER} class from a Dining Philosophers solution in \scoop~\cite{Referee-Webpage}}
  \begin{lstlisting}[basicstyle=\scriptsize]
live -- Each Philosopher eats times_to_eat times
  do
    from until
      times_to_eat < 1
    loop
      print ("Philosopher " + Current.id.out  + " waiting for forks.%N")
      eat (left_fork, right_fork)
      print ("Philosopher " + Current.id.out  + " has eaten.%N")
      times_to_eat := times_to_eat - 1
    end
  end (*@\\[-2ex]@*)
eat (left, right: separate FORK) -- Eat, having acquired left and right forks
  do
    -- Here, eating takes place
  end(*@\\[-2ex]@*)
left_fork, right_fork: separate FORK -- References to forks used for eating
\end{lstlisting}
\label{lst:dining_philosophers}
\end{Listing}

\paragraph{Concurrency, Asynchronicity, and Locking}
\scoop has a number of other features and behaviours detailed more thoroughly in \cite{Morandi-SNM13a}; here, we briefly describe only queries and contracts. First, given a separate object reference \lstinline{x} and a method \lstinline{m} that is a \emph{query} (i.e.\ returns a result), a call \lstinline{q = x.m} is always executed synchronously; if \lstinline{x} has a separate handler, then the current object's handler simply waits for the result to be returned. Secondly, \scoop maintains and extends the Eiffel tradition of annotating methods with preconditions (keyword \eifblue{require}) and postconditions (\eifblue{ensure}). In the sequential setting, these are (optionally) checked before and after every execution of the method. In the concurrent setting, however, preconditions are instead interpreted as \emph{wait conditions} that must be synchronised on. Conceptually, the execution of a routine is delayed until simultaneously the precondition is satisfied and the handlers of the formal arguments are locked. This allows the programmer to express synchronisation conditions at a high level of abstraction.

These concepts require execution-time support from an effective run-time environment. The current run-time \cite{Morandi-SNM13a} associates each processor with a lock and \emph{request queue}. A method call on a separate object is enqueued on its handler's request queue, which is processed in FIFO order. The call may only be enqueued if the lock on the handler is held. The run-time is responsible for correctly synchronising on wait conditions and locking the handlers of formal arguments at the beginning of methods, as well as releasing them at the end. The run-time must balance these design needs with the need to permit a reasonable level of performance (e.g.~by reducing resource contentions). As such, and as a major challenge for our work, the ``official'' run-time is frequently evolving, and several alternatives have been proposed and/or developed, e.g.\ \cite{Nienaltowski07a,Morandi-SNM13a,West-NM15a}.


\section{A \gts-based Model of \corescoop}
\label{sec:cpm}

Our first step is to present a run-time model for the core behaviours of \scoop, i.e.\ remote method calls, FIFO queues, and locking. This model, named \emph{Concurrent Processor Model} (\cpm), strips away the object-oriented features of \scoop, grasping only a subset of the language and focusing on processors equipped with simple data. This allows us to:
\begin{inparaenum}
  \item highlight the fundamental peculiarities of \scoop as model of concurrency in a more fine-grained formal setting, and
  \item present the basic building blocks of our approach in more detail, as we extend \cpm to include full object-orientation in Section~\ref{sec:fullscoop}.
\end{inparaenum}

\begin{figure}[t!]
 \centering
 \includegraphics{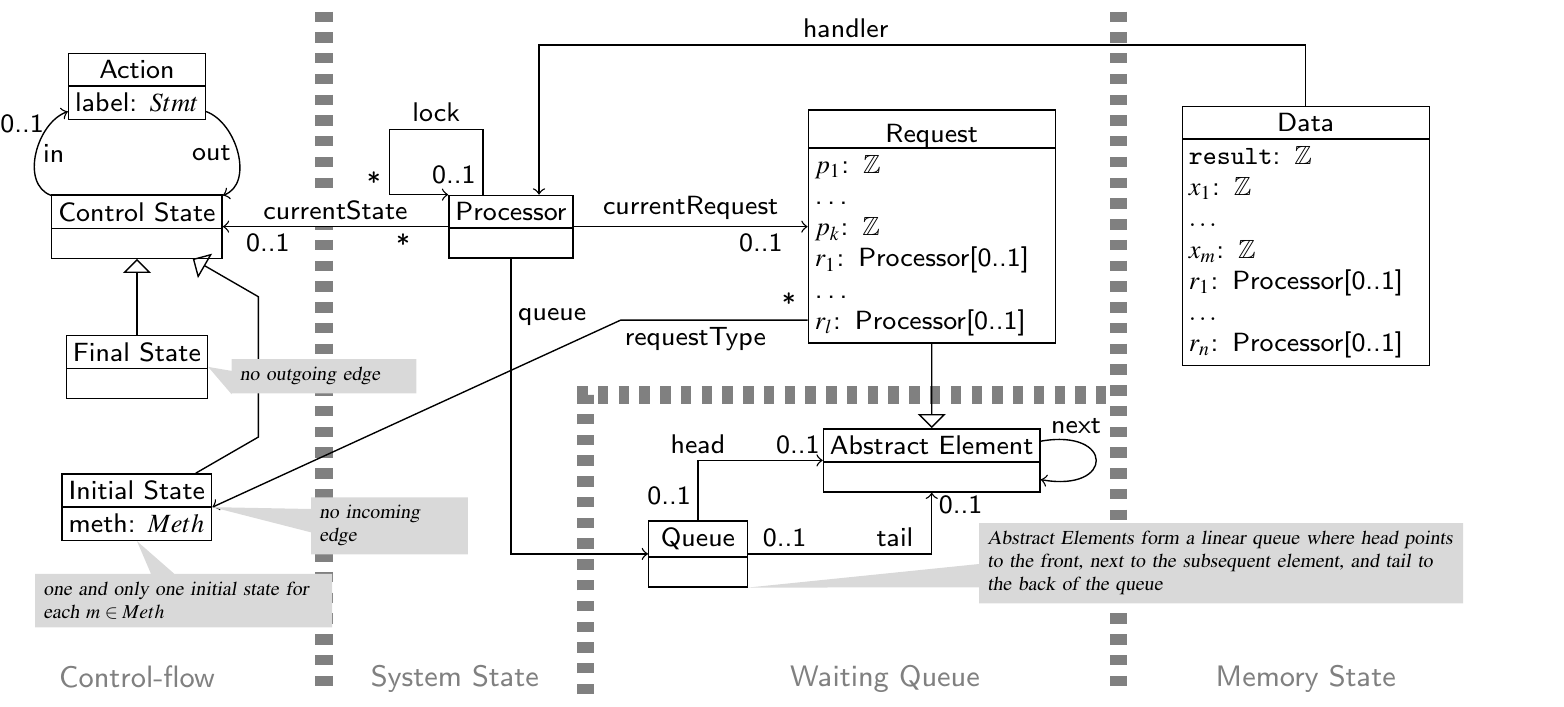}
 \caption{Type graph for \cpm's configurations as class diagram with constraints, where we
 assume disjoint finite sets of variable names ($p_1,\dots, x_1\dots$), and reference names ($r_1,\dots$); let \Meth be a finite set of method names, and $\ZZ$ be the set of integer numbers, \BB of Booleans; the cardinality of association edges is "1" if not noted otherwise; the different regions highlight the different modules of the model
 }\label{fig:cpmtypegraph}
\end{figure}

\paragraph{From \corescoop to \cpm}
Stripping ``local'' object-oriented features from \scoop (e.g.\ self-calls, non-separate calls) and focusing on \emph{remote} synchronous and asynchronous method calls (i.e.\ queries and commands) via FIFO queues, as well as locking in a concurrent setting, leads to a subset of the \scoop language we call \corescoop in the following.
We formalise the run-time model for \corescoop by the \emph{Concurrent Processor Model} (\cpm). \cpm is represented by a graph transformation system in which configurations are given by directed graphs conforming to the type graph of Figure~\ref{fig:cpmtypegraph}. Note that the type graph uses a UML-like notation with type attributes and constraints.

The semantic model of \corescoop is inspired by the current ``standard'' formalisation of \scoop's semantics in~\cite{Morandi-SNM13a}. It consists of a set of \emph{processors} that run concurrently. Each processor is the \emph{handler} of data in local memory, which is represented as a mapping from variable names ($x_1,\dots x_m$) to integers. There is \emph{no global shared memory}, only processor local memory, and this memory can only be accessed by or via its processor.
Processors sequentially handle method calls via incoming \emph{requests} that are related to a control-flow graph encoding of the underlying \corescoop methods. Thus, a running processor that handles a \emph{current request} is in a \emph{current state} belonging to this \emph{request method type's} control flow graph. Incoming requests are stored by each processor in a FIFO \emph{queue} before being locally executed.
Each processor has a finite set of known neighbour processors, i.e.~those accessible for synchronous or asynchronous remote calls, which are stored by reference (the variables $r_1,\dots,r_n$).
Processors can dynamically generate new processors (and assign these directly to local reference variables).
Each remote call and its context, i.e.~the call's parameters, which consist of integer values (i.e. $p_1,\dots p_k$) and processor references ($r_1,\dots, r_l$), is stored as a \emph{request}. Requests implement ``value passing'', e.g.~requests can pass references to newly generated processors.
The return value for queries, i.e. synchronous remote calls, is stored in a special local variable accessible to the caller (variable \lstinline{result}). In \cpm, there are neither local calls, i.e.~calls to oneself, nor local recursion.

\cpm includes {explicit locking} between processors, i.e.\ each processor can be locked by at most one (distinct) processor. \corescoop's implicit locking can thus be translated to \cpm's explicit locking, however at a different level of granularity.
In general, \cpm is able to \emph{simulate} the execution of programs written in \corescoop, which is not possible the other way round due to this different level of atomicity in both models.
%
%
A processor handling a request ``walks'' the corresponding part of the control-flow graph by updating the current state according to the actions' semantics, given as graph transformation rules (see Section~\ref{sec:withgroove} for example configurations and a rule for commands). Handling of queues, local scheduling of each processor (i.e.~terminating the currently handled request and advancing to the next in the waiting queue), instantiation of parameters, etc., is handled by global \emph{scheduling rules}, which also assure that each local processor advances as far as possible. Note that the walking of the control-flow graph and the scheduler are generic, i.e.~represented by a set of \gts rules independent of the \scoop program. We refer to \onlinerepo for the full formal model directly represented as \gts (that can be browsed and simulated with \groove).

\paragraph{Modularity of \cpm}
Configurations of \cpm, i.e.\ global states of the system, can be partitioned into four main parts, which are also visible in Figure~\ref{fig:cpmtypegraph}:
\begin{inparaenum}
\item a representation of the underlying \scoop program in the form of a control-flow graph;
\item a system of concurrently running processors, each one possibly handling a request; these processors represent the \emph{system state};
\item a waiting queue for each processor that stores pending requests;
\item local \emph{memory state} for each processor.
\end{inparaenum}
The control flow component can be derived directly from the original \corescoop program's control flow graph (at a pre-compilation step) and its structure does not dynamically change, contrary to the state of the run-time environment (processors, queues, data).
This partitioning is also mirrored by the \gts's underlying rules that treat the walking of the control flow graph separately from the queue's policies, the management of the memory, and global scheduling.

The fixed simple interfaces (in form of the model components' loose coupling due to the typed associations {\sf queue}, {\sf handler}, {\sf currentState}, and {\sf requestType}) between these modules allow us to \emph{plug in different behaviours} for each module, e.g.\ different queueing semantics. Thus we can either adapt different existing \scoop semantics (e.g.~FIFO queues versus queues of queues~\cite{West-NM15a}) or directly apply abstraction mechanisms in the context of verification (e.g.~a counting abstraction of the queue's content, or predicate abstraction for the data) by small modular changes to the underlying rules.

Furthermore, the global abstract scheduling rules can be parameterised in this way, e.g.\ to include different kinds of garbage collection in the global scheduler or different rule prioritisations that keep the state space small, such as always preferring to terminate processors that are currently in a final state.


\section{Simulating CPM in \groove}
\label{sec:withgroove}

We realised \cpm---our run-time model for \corescoop---in GROOVE, an established tool for simulating and analysing GTS-based semantics. This section describes how we approached and achieved this task. First, we justify our choice of GROOVE, and then show (by example) how \cpm configurations, rules, and rule applications are represented in the tool. Finally, we discuss the issue of \cpm's soundness.

\paragraph{The \gts Tool \groove}
We chose \emph{GRaphs for Object-Oriented VErification} (\groove)~\cite{GROOVE-Reference,Ghamarian-MRZZ12a} as our platform to implement and analyse the \cpm models. Most existing \gts tools are in theory expressive enough to cover \cpm. \groove however was already applied for the analysis of (non-concurrent) object-oriented programs in Java~\cite{RensinkZ09}. Furthermore, \groove contains a (finite-state) model checker that has proven sufficient for the analysis and verification of dynamic state systems~\cite{DelzannoT13,KastenbergR06}. As reported in \cite{Zambon-Rensink14a}, GROOVE can typically handle systems with up to 4 million states, which should leave enough room for our first experiments. Finally, \groove convinced us with a gentle learning curve, its ease of adaption and extension to our needs, as well as its active development community. 

\paragraph{Representing \cpm Configurations in \groove}

\cpm configurations are represented in GROOVE quite straightforwardly, with control-flow, system state, waiting queue(s), and memory state (as in the type graph of Figure \ref{fig:cpmtypegraph}) all encoded in the same graph.

\begin{wrapfigure}{r}{0.6\textwidth} 
  \begin{center}
  \vspace{-10pt}
    \includegraphics[scale=0.67]{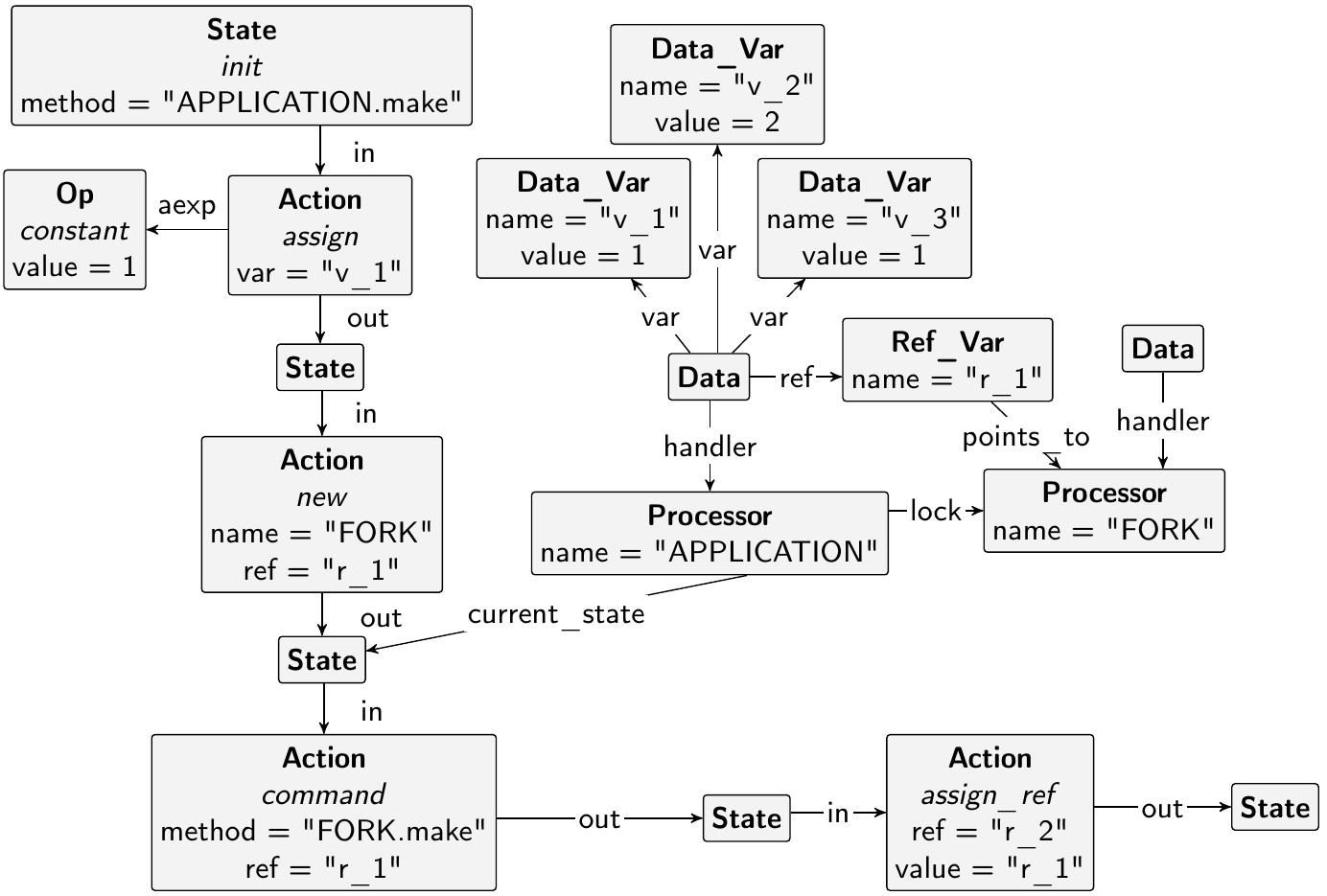} 
  \end{center}
  \caption{A \cpm configuration rendered in \groove}
  \vspace{-10pt}
  \label{fig:cpmconfiguration}
\end{wrapfigure}

Control-flow is rendered as static transition systems in the graph. These comprise State nodes, where entry points are labelled with \emph{init} and a $\tt method$ name, and exit points with \emph{final}. A transition between two State nodes is encoded as a pair of edges ($\tt in$ and $\tt out$) and an Action node labelled with a \cpm action (e.g.\ \emph{command}, \emph{query}, \emph{lock}, \emph{assign}). Encoding actions as nodes---as opposed to labelled edges between states---facilitates a clean way of modelling \emph{action parameters}. Simple action parameters, such as a variable to assign to, are encoded as attributes of Action nodes; compound action parameters on the other hand, such as a Boolean expression to be evaluated, are modelled as abstract syntax trees incident to Action nodes. Furthermore, for actions that trigger methods on other processors (i.e.\ commands, queries), an arbitrary number of \emph{method parameter} nodes (which represent data to be instantiated and available for the duration of a method) can be attached to the corresponding Action nodes. These encode, via attributes, the parameter name, as well as the integer or reference variable to instantiate and pass as the method parameter.

System state, waiting queue(s), and memory state are rendered as dynamic parts of the graph. Each Processor node handles a Data node. Data nodes are incident to Data\_Var and Ref\_Var nodes, which respectively store the handling Processor's integer variables (via attributes) and reference variables (via edges). Processors may hold locks on other Processors (represented by $\tt lock$ edges), and may be in a control State (represented by a $\tt current\_state$ edge). Furthermore, they have waiting queues of requests to be executed, represented as ``linked lists'' of Queue\_Item nodes over $\tt next$-labelled edges. Each such node is labelled with the $\tt method$ (i.e.\ the particular transition system) to be executed, and is attached to nodes that store the values of any method parameters expected.

An example of a \cpm configuration in \groove is given in Figure \ref{fig:cpmconfiguration}. This shows a subgraph of a configuration that can be reached in the \cpm encoding of the full Dining Philosophers SCOOP program (see \cite{Referee-Webpage}). There are two Processors in this configuration, but only one (APPLICATION) is currently executing a method ($\tt current\_state$). This Processor has both a reference to ($\tt points\_to$) and a $\tt lock$ on the FORK processor (i.e.\ it has the exclusive right to send requests). Neither Processor is storing any method parameters, and their waiting queues are empty.

\paragraph{Simulating \cpm Actions in \groove}

The semantics of \cpm is simulated in \groove by two sets of graph transformation rules: action rules, and scheduling rules. 

Action rules facilitate the firing of transitions in the control-flow part of the graph, and the corresponding updates to the system and memory state. They model the basic behaviours of \scoop processors: variable assignment, condition evaluation, processor creation, asynchronous commands, synchronous queries, and simultaneously (un)locking multiple processors. An action rule can be applied to a \cpm configuration when:
\begin{inparaenum}
\item a processor is in a control-flow state incident to a correspondingly-labelled action; and
\item the prerequisites of the action are satisfied, e.g.\ every processor targeted by a lock action is available to be locked.
\end{inparaenum}
Action rules are atomic, in that the firing of a transition occurs in a single, indivisible step (e.g.\ locking multiple processors occurs instantly, as it appears to \scoop programmers). This is achieved by extensive use of \groove's powerful matching constructs such as \emph{universal quantification}, which allows for a single rule to handle arbitrarily many instances of particular sub-structures (e.g.\ arbitrary numbers of method parameters). Furthermore, action rules are assigned the same (and lowest) priority in \groove, meaning that non-determinism (and thus interleaving) is modelled at the level of atomic processor actions, as opposed to partial evaluations (thereby mitigating one source of state space explosion).
%
\begin{figure}[t!]
	\centering
	\includegraphics{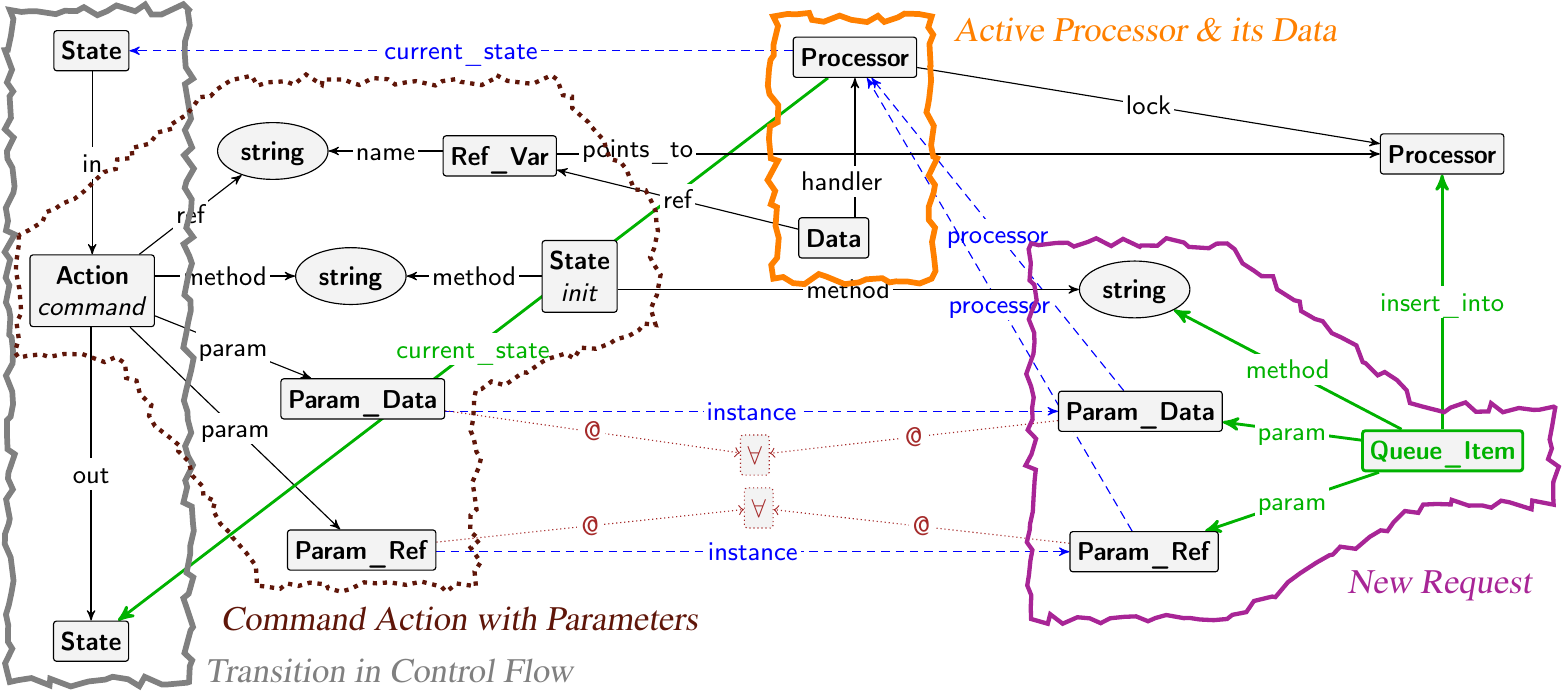}
  \caption{Example action rule for commands in \groove (\groove's rule colour coding: \blue{dashed blue} elements only exist on the left-hand side of rule (thus will be deleted), \green{bold green} elements on the right-hand side (thus will be generated), black ones persist)}\label{fig:examplerule}
\end{figure}

Scheduling rules handle queues, local scheduling of each processor (e.g.\ advancing to the next request), and any local pre- or post-processing required for action rules; more generally, they advance processors as much as possible in-between actions. While action rules necessarily model non-determinism---different interleavings model different orders in which requests are enqueued, and thus potentially different program outcomes---scheduling rules avoid it as much as possible, since the steps between actions are \emph{local} to processors. This is achieved by rule priorities in \groove. In particular, all scheduling rules have higher priorities than action rules, meaning that all local scheduling is simulated before exploring the non-determinism of actions. Furthermore, no two scheduling rules have the same priority, ensuring that their execution is as deterministic as possible to reduce the number of states to explore. Assigning such fine-grained rule priorities did, however, require some care. It is ultimately unimportant, for example, whether a constant or variable in an expression is evaluated first, so we arbitrarily fixed the priority of the scheduling rule for constants to be higher. On the other hand, if we had assigned the scheduling rule for terminating requests (i.e.\ in final states) to be of higher priority than scheduling rules that perform post-processing immediately after actions (e.g.\ ``resetting'' an evaluated expression after assigning it), then a fault would have been introduced into the model.

Let us take a closer look at the action rule for commands, depicted in Figure \ref{fig:examplerule} using \groove's colour coding. Recall that in \scoop (and thus in \cpm), a command is an asynchronous remote method call on a concurrently running processor that is locked by the current processor.
The starting point of the rule is an active processor that is currently in a state (\gblue{current\_state}) that could fire a command action. The command action is given by the following:
\begin{inparaenum}
\item a pointer ($\tt points\_to$) from a reference variable (Ref\_Var) to the target Processor;
\item the name of the $\tt method$ to call, via an attribute of the Action node; and
\item some number of instantiated method parameters (Param\_Data, Param\_Ref).
\end{inparaenum}
The firing of the rule advances the processor's current state (\ggreen{current\_state}) and at the same time generates a new request
\begin{wrapfigure}{r}{0.6\textwidth} 
  \begin{center}
  \vspace{-15pt}
    \includegraphics[scale=0.67]{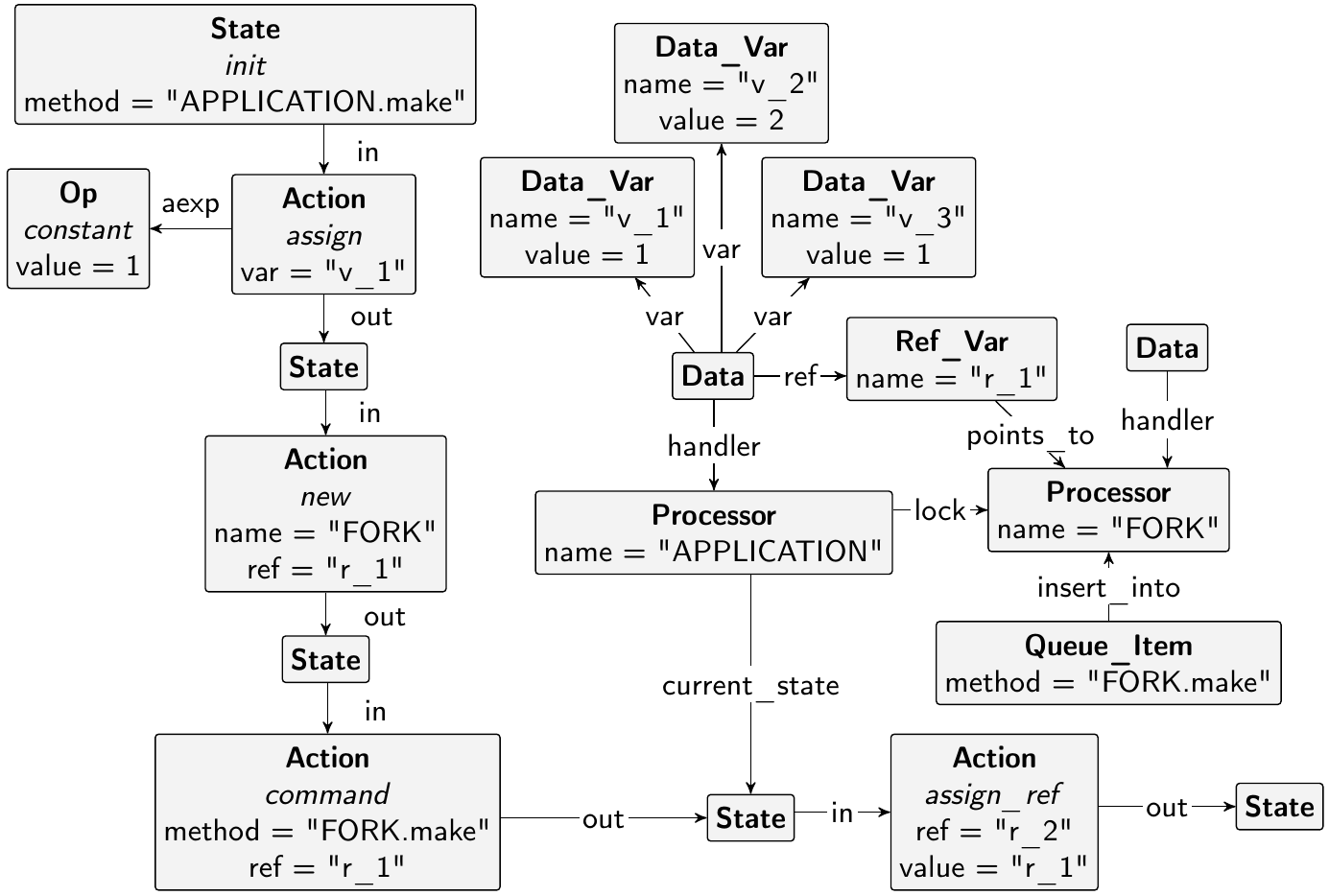} 
  \end{center}
  \caption{Effect of the command rule on Figure \ref{fig:cpmconfiguration}}
  \vspace{-10pt}
  \label{fig:cpmruleapplication}
\end{wrapfigure}
(as \ggreen{Queue\_Item}) equipped with the $\tt method$ name and all (via ``\textcolor{red}{$\forall$}'') instantiated parameters. This is then later inserted into the target Processor's waiting queue via a separate, highly prioritised scheduling rule. For illustration, the result of applying the action rule to Figure \ref{fig:cpmconfiguration} is given in Figure \ref{fig:cpmruleapplication}. Observe that the transition has been fired, that the resulting Queue\_Item contains the same $\tt method$ name as the Action node, and that it is waiting to be enqueued by the Processor that the reference variable $\tt r\_1$ points to.
The current prototype of \cpm is available at \cite{Referee-Webpage}. At present, it comprises 19 action rules and 34 scheduling rules.

\paragraph{Soundness}

Formally establishing the soundness (or ``faithfulness'') of \cpm is an important step, which would rely on either the existence of a definitive language semantics, or an in-depth comparison with one of the proposed semantic approaches to \scoop. Since the former is lacking, we have attempted to be faithful to the comprehensive operational semantics proposed by Morandi et al.\ \cite{Morandi-SNM13a} which is ``executable'' in Maude. To prove soundness with respect to this semantics would however require a more stringent formal representation of \cpm, e.g.\ as graph programs \cite{Poskitt-Plump14a}, and corresponding semantic proofs (e.g. equivalence/(bi-)simulation). We are addressing this as ongoing work, but it is beyond the scope of this first, proof-of-concept paper.

In the meantime, we took a more lightweight approach to gain confidence in the soundness of \cpm, through example-driven testing and an expert review. For the former, we looked at the \scoop examples supplied with the EiffelStudio IDE, which demonstrate idiomatic usage of \scoop's concurrency mechanisms to solve a number of classical synchronisation problems. We focused on two programs in particular---Dining Philosophers and Single-Element Producer Consumer---which only ever create one object per processor, and thus were \corescoop programs that straightforwardly map to \cpm actions (recall that \cpm does not model the notion of multiple local objects). With these programs, we then tested the faithfulness of \cpm by:
\begin{inparaenum}
\item visualising and manually inspecting program executions in the \groove simulator;
\item exploring the state space for abnormal states (e.g.\ unsatisfied action pre-requisites, such as the absence of a lock edge for a command) using the LTL model checker (see Section \ref{sec:verif}); and
\item comparing the effects of action rules against the informal and formal descriptions of \scoop in \cite{Morandi-SNM13a}.
\end{inparaenum}
In addition to testing, we also held a one-day ``expert review'' with B.~Morandi and other \scoop researchers from ETH, during which we demonstrated and discussed the \cpm rules in detail with the goal to ensure that the rules fully corresponded to their understanding of \scoop/\corescoop.


\section{Towards Full-Fledged \scoop, Approximations, and Translations}
\label{sec:fullscoop}

In this section, we look beyond \cpm and \corescoop to consider three ongoing extensions to the work. First, we describe our effort to extend the model with full object-orientation, and thus make it expressive enough for a wider class of \scoop programs. Second, we discuss how \cpm can be used as a basis for exploring alternative \scoop semantics and model approximations. Finally, we report on a prototype tool for automatically generating \groove input from \scoop (and thus also \corescoop) programs.

\paragraph{From \cpm to \cpmoo} 
\cpm allowed us to ``boil down'' \scoop to the core of its asynchronous, concurrent behaviour, and study it in a formal setting without the full complexity of object-orientation. Our aim however is practical verification, and in practice, \scoop programmers extensively use objects: ultimately we need to support them. The benefits of a simpler formal model aside, one might wonder why we did not start with full object-orientation from the outset if practical verification was always in mind. This is because it allowed a separation of concerns: we could first isolate and model concurrency-via-processors in a clean, simple setting, and then separately extend it with the object-oriented and Eiffel-specific languages features that are not core to the concurrency model. Our modelling approach has essentially been to identify this core, formalise it, then gradually add the missing details and behaviours.

We are extending \cpm to \emph{CPM with Object-Orientation} (\cpmoo), a richer run-time model capable of expressing and simulating \scoop programs with multiple objects per processor and non-separate method calls (i.e.\ targetting local objects). The present version of \cpmoo is the result of the following process:
\begin{inparaenum}
\item replacing simple data in the \cpm type graph with objects;
\item updating the rules that then no longer conform, in consultation with the semantics of \cite{Morandi-SNM13a}; and
\item testing (including regression testing for \corescoop programs).
\end{inparaenum}

Our first goal was to support all of the \emph{existing} actions of \cpm, but with data organised into objects. We began by updating the type graph, replacing simple data with object nodes connected to attributes; attributes being integers (as before) or references to other \emph{objects} (not processors). The advantage in changing the type graph first is the instant feedback from \groove, which highlights the rules that no longer conform and thus need updating (i.e.\ every rule that processed or extracted data). In general, these were not radical updates: the core behaviour captured in \cpm remained the same, and the semantics of actions did not fundamentally change. What had to be updated was the structure of data that sat on top of this core, as well as remote calls to processors which became remote calls to objects. In other words, the question we were asking at each step was ``how do we correctly embed objects into the semantics of this action'' and emphatically \emph{not} ``how do we model this asynchronous behaviour for objects''.

With \cpm ``objectified'', we could turn to modelling behaviours only possible with data organised into objects, most notably, non-separate calls (calls to objects on the same processor). There is of course no reason to acquire locks in the non-separate case, and the processor simply executes the method immediately. To model this, we had to first model the call stack, also allowing us to capture recursion and local variables---important, practical details, but ultimately on top of (and not crucial to) the concurrency at the core. The present prototype of \cpmoo, available to download from \cite{Referee-Webpage}, includes all of the features discussed, as well as arbitrary names for attributes (e.g.\ $\tt buffer$ instead of $\tt r\_1$), separate queries in expressions, reference expressions, and (optional) postcondition checking.

To gain confidence that the extension to \cpmoo remained sound, we followed a similar testing approach to that described in Section \ref{sec:withgroove}, but using a wider selection of the example \scoop programs distributed with EiffelStudio (since the model is now expressive enough to simulate them). In addition, we also:
\begin{inparaenum}
\item used a number of simple sequential programs (i.e.\ \scoop programs with only one processor) to focus some testing on the new rules for non-separate calls; and
\item performed ``regression testing'', in the sense of ensuring that \corescoop programs do not behave differently in \cpmoo to basic \cpm.
\end{inparaenum}

\cpmoo does not yet cover all of \scoop: many of the Eiffel-specific mechanisms (agents, once routines, exceptions) and their interactions with \scoop have not been captured, nor have some advanced run-time mechanisms such as separate callbacks. We also have ignored inheritance for the moment (following \cite{Morandi-SNM13a}), viewing it as an advanced typing mechanism and a separate problem for us to tackle. We do not anticipate substantial difficulty in adding them to \cpmoo; it is our plan to eventually include them by the same methodology, which we view as a promising, practical means of facilitating verification for a rich, complex concurrency model like \scoop.

\paragraph{Run-Time Alternatives and Approximations}

An alternative to this gradual extension of \cpm is to use it as a basis for exploring and prototyping \emph{alternative semantics}. For \scoop this is particularly important, since the model has so many competing semantics; most recently a proposal to replace each FIFO queue with a queue of queues \cite{West-NM15a}. Changing the \gts implementation of the waiting queue, for example, is relatively straightforward due to the model's modularity (see Section \ref{sec:cpm}). Rather than changing the \scoop compiler first, and risking the discovery of fundamental problems after having committed the time, we propose modifying \cpm first, comparing execution traces, and ensuring that the changes retain the high-level guarantees of the model and any other desired properties. We are exploring this usage of \cpm as ongoing work, but envisage that such prototyping can be achieved in an analogous way to adding object support: modify the type graph first (e.g.\ replace the FIFO queue with a queue of queues), revise the affected rules, and test.

A similar idea is to implement \emph{approximations} of \cpm directly in the \gts, by plugging in different scheduling rules. As an example, we replaced the FIFO queues of processors with bags (see \cite{Referee-Webpage}), thereby removing the guarantee of processors executing their requests in the order they were received. This is an over-approximation in the sense that all the behaviours of FIFO queues are included in the state space, but several other infeasible behaviours are included too (hence verification of the approximation implies verification of the program, but a counterexample may simply be spurious). Going further, we could, for example, over-approximate \cpm by a Petri net (also represented as a \gts).

\paragraph{Translating \scoop Programs to \groove}
We are developing a tool (to be published as part of a Master's thesis) that automatically translates \scoop programs to input graphs for \groove. The tool targets the same subset of the \scoop language that \cpmoo prototype formalises (it completely handles, for example, all of the \scoop programs in \cite{Referee-Webpage}). The current prototype first creates a syntax tree of the input classes using the ANTLR4 parser generator in conjunction with an existing \scoop grammar. Since the input graph requires some typing information (for example, there are different action nodes for integer and reference assignment), the tool passes through the syntax tree twice; first to gather typing information, and then again to generate an intermediate representation of the program that is closely related to the input graph. Finally, the tool passes through the intermediate representation and generates the corresponding input graph as an XML file conforming to the Graph eXchange Language. This graph can then be interpreted and analysed by \groove.


\section{Verification of \scoop Programs}
\label{sec:verif}

In this section we explore how a \scoop program, once translated to our run-time model in \groove, can be verified by (bounded) model checking. After discussing the kinds of properties that can be checked, we illustrate the detection of deadlock in a faulty Dining Philosophers \scoop solution, and obtain some first verification impressions in a small evaluation of five \scoop programs.

\paragraph{Verification}

The \groove model checker can be used for automatic analyses that are based on the idea of determining the presence (or absence) of a state that violates some expected property of the program. One such property---the absence of deadlocks---provided the initial motivation for this work. The range of properties that can be verified, however, is much broader; two contrasting examples include the absence of calls to void (null) object references, and the absence of states that violate postconditions (see \cite{Referee-Webpage}). This is achieved by extending the run-time model with a set of error rules (assigned the highest priority) that match if and only if the current configuration violates a particular property. An error rule for deadlock, for example, will match if there is a cycle of processors in states preceding lock actions, such that each lock action requires, in turn, a resource held by the next in the cycle. To catch a void call, on the other hand, an error rule will match if a processor is in a state immediately before an action that targets a void reference variable. Then, verification by model checking is simply a matter of expressing---in a temporal logic formula over rules---that none of these error rules are ever applied in the state space. For programs that have an infinite state space (our examples here do not, but those derived from general \scoop programs may), \groove supports \emph{bounded} model checking, which, although unable to fully guarantee correctness, does provide a means of searching for the presence of counterexamples. See~\cite{DelzannoT13} for details on bounded model checking with \groove.

Recall the Dining Philosophers example from Listing \ref{lst:dining_philosophers}. This implementation avoids the possibility of deadlock because the $\tt eat$ method requires the simultaneous acquisition of locks on the handlers of the forks (in \cpm, this implicit locking is expressed in a single action). Suppose that the philosophers instead call $\tt bad\_eat$, as given in Listing \ref{lst:dining_philosophers_bad}. This implementation permits executions that lead to deadlock, since philosophers now pick up their forks in turn (which in \cpm, then maps to two distinct locking actions).
\begin{wrapfigure}{r}{0.5\textwidth}
  \begin{center}
  \vspace{-7pt}
    \includegraphics[width=\linewidth]{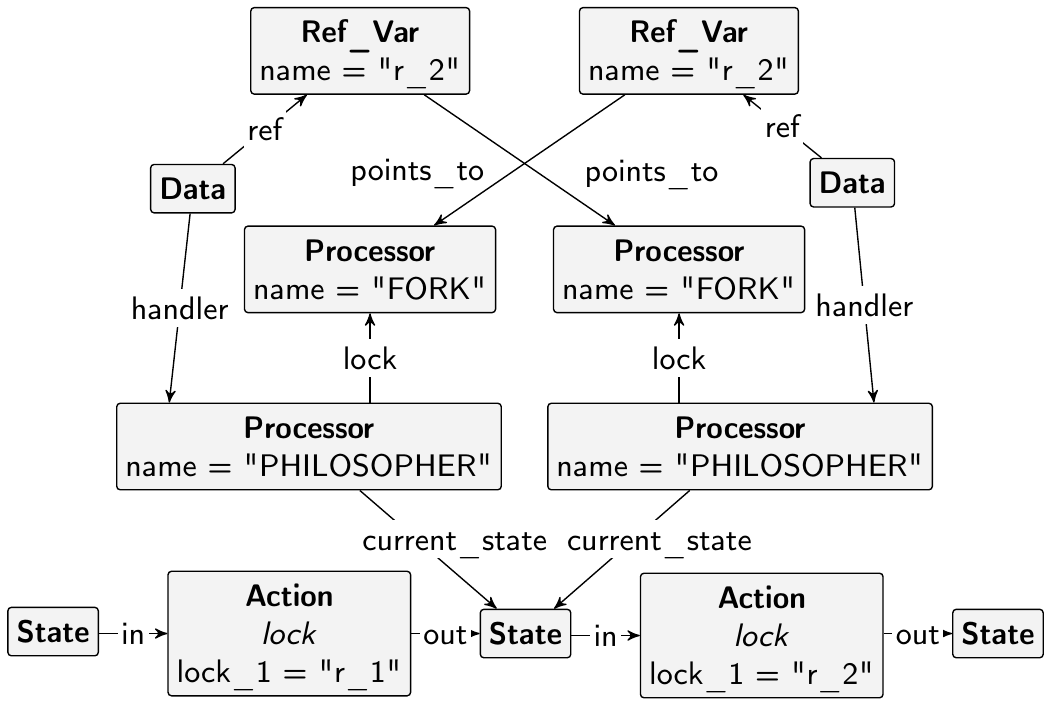}
  \end{center}
  \caption{A deadlocked \cpm configuration}
  \vspace{-10pt}
  \label{fig:deadlock}
\end{wrapfigure}
In particular, if every philosopher locks the handlers of their left forks by reaching line 8, the system will deadlock since every fork is locked, preventing the philosophers from entering $\tt pickup\_right$. Using the error rules for deadlock and the model checker of \groove, faulty executions are automatically unearthed and reported, i.e.\ paths through the state space from the initial configuration to states exhibiting the structural ``deadlock pattern''. The relevant part of such a deadlocked state for two philosophers is given in Figure \ref{fig:deadlock}. Here, the philosophers have already locked their left forks, and both are waiting to lock their right ones (on reference variables $\tt r\_2$). Since the right fork of each philosopher is the already-locked left fork of the other, neither processor can fire the action, and the system is deadlocked.

\begin{Listing}[!tb]
  \caption{Snippet of a Dining Philosophers implementation that may deadlock}
  \begin{lstlisting}[basicstyle=\scriptsize]
bad_eat
  do
    print ("Philosopher " + Current.id.out  + " waiting for left fork.%N")
    pickup_left_then_right (left_fork)
  end
pickup_left_then_right (left: separate FORK)
  do
    print ("Philosopher " + Current.id.out  + " waiting for right fork.%N")
    pickup_right (right_fork)
  end
pickup_right (right: separate FORK)
  do
    -- Here, eating takes place
  end
\end{lstlisting}\vspace{-15pt}
\label{lst:dining_philosophers_bad}
\end{Listing}

While this particular bug is somewhat contrived, it does illustrate a discord between the programmer's level of abstraction---``here are the concurrent objects that my method needs''---and a run-time that attempts to handle it all under the hood, but can ultimately fail if the programmer ignores (or is unaware of) how it works. Beyond this example, there are more subtle ways in which deadlock can unintentionally and accidentally be introduced in \scoop programs~\cite{West-NM10a}.

\paragraph{Evaluation}

To obtain some initial impressions of verification performance, we ran a small study on the current \cpmoo prototype. We devised ten benchmarks from various configurations of five \scoop programs: Dining Philosophers, both a version that calls $\tt eat$ (DPE) and another that calls $\tt bad\_eat$ (DPB), single-element Producer Consumer (PC), Dining Savages (DS), and Cigarette Smokers (CS). These are available at \cite{Referee-Webpage} and were adapted (i.e.\ to replace unsupported features like inheritance) from the example \scoop programs provided with EiffelStudio (except for CS, which we implemented).

First, we used the current version of our translation tool to generate input graphs. Then, for these programs, we recorded data about graph sizes, the time to check for deadlock, the time to explore the full (finite) state spaces, and peak memory usage.  Table~\ref{tab:performance} presents this information, computed on an off-the-shelf workstation with Intel Core i7-4810MQ CPU and 16 GB main memory. The presented values are the medians of five tests. Here, $n$ refers to the number of philosophers (for DPE and DPB), the number of elements to produce (for PC), or, for DS, respectively the pot size, number of savages, and hunger. Time was measured using System.nanoTime(); the results represent the elapsed (wall clock) time in seconds. Memory usage was measured using Java's MemoryPoolMXBean and related classes. Finally, the exploration strategies used were $\tt bfs / final / infinite$ and $\tt ltl(prop=!F\ error\_deadlock) / final / infinite$ for the full state space and LTL exploration respectively.

The graph sizes differ little between initial and final states, with the only variation due to the creation and manipulation of processors, their waiting queues, and objects in the local memory. Note the performance discrepancy between LTL checking and full state space exploration. For DPB, which may deadlock, LTL checking is faster since finding one counterexample is enough to return an answer. For all the other programs, which do not deadlock, checking the formula incurs an overhead. Across most of the benchmarks, we would argue that the times are acceptable and practical (especially given the infeasibility of model checking the Maude semantics \cite{Morandi-SNM13a}). An exception is DS, where the overhead for LTL checking is substantial for $n =(2,4,1)$. Understanding the reasons for this is part of an ongoing, broader investigation into the scalability and limits of the tool for verifying \scoop programs.

\begin{table}[tb]
	\caption{First impression of verification performance}
	\label{tab:performance}
	\centering
	\scriptsize
		\begin{tabular}{|c|c|c|c|c|c|c|c|c|}
		\hline
		Program ($n$) & Start Graph & Final Graph & \multicolumn{2}{c|}{LTL Deadlock} & \multicolumn{3}{c|}{Full state space} \\
		\cline{4-5}\cline{6-8}
		  & (nodes/edges) & (nodes/edges) & time (s) & (states/transitions) & time (s) & (states/transitions) & Mem.\ [stddev] (GB) \\
		\hline\hline
		DPE (2) & 326/496 & 362/582 & 1.10 & 824/838 & 1.18 & 824/838 & 0.65 [0.11] \\
		DPE (5) & 326/496 & 389/653 & 32.37 & 20,428/20,906 & 29.10 & 20,428/20,906 & 4.23 [0.98] \\
		\hline
		DPB (2) & 322/488 & 378/644 & 0.84 & 708/712 & 1.33 & 1,108/1,134 & 0.55 [0.09] \\
		DPB (5) & 322/488 & 447/836 & 175 & 74,942/77,378 & 204 & 122,714/127,425 & 5.62 [0.20] \\
		\hline
		PC (5) & 371/563 & 393/621 & 3.51 & 2,152/2,194 & 3.30 & 2,152/2,194 & 0.64 [0.14] \\
		PC (20) & 371/563 & 393/621 & 12.98 & 8,362/8,539 & 12.84 & 8,362/8,539 & 1.42 [0.24] \\
		\hline
		DS (2,2,2) & 440/668 & 470/749 & 11.48 & 5,976/6,081 & 10.82 & 5,976/6,081 & 1.41 [0.32] \\
		DS (2,3,2) & 441/668 & 478/769 & 388 & 103,190/106,260 & 256 & 103,190/106,260 & 5.71 [0.29] \\
		DS (2,4,1) & 441/668 & 486/789 & 2,448 & 396,011/414,462 & 941 & 306,401/319,018 & 7.28 [0.63] \\
		\hline
		CS & 559/866 & 608/1000 & 417 & 65,275/70,008 & 370 & 65,275/70,008 & 5.27 [0.23] \\
		\hline
	\end{tabular}
\end{table}


\section{Conclusion}
\label{sec:conclusion}

\paragraph{Related Approaches for \gts-based Specification and Analysis of Concurrent OO Programs}
Verifying concurrent object-oriented programs with \gts-based models is an emerging trend in software specification and analysis, especially for approaches rooted in practice. See \cite{Rensink10} for a good overview discussion, based on a lot of personal experience, on the general appropriateness of \gts for this task.

Closest to our semantic run-time model is the QDAS model presented in \cite{Geeraerts-HR13a}, which represents an asynchronous, concurrent waiting queue based model with \emph{global memory} as \gts, for verifying programs written in Grand Central Dispatch. Despite the formalisation as \gts, there is, however, no direct compiler to \gts yet. The Creol model of \cite{JohnsenOY06} focuses on asynchronous concurrent models but without more advanced remote calls via queues as needed for \scoop. Analysis of the model can be done via an implementation in a term rewriting tool~\cite{JohnsenOA05}.
Existing \gts-based models for Java only translate the code to a typed graph similar to the control-flow sub-graph of \cpm \cite{RensinkZ09,CorradiniDFR04}. A different approach is taken by \cite{FerreiraFR07}, which abstracts a \gts-based model for concurrent OO systems~\cite{FerreiraR05}  to a finite state model that can be verified using the SPIN model checker.
\groove itself was already used for verifying concurrent distributed algorithms on an abstract \gts level~\cite{Ghamarian-MRZZ12a}, but not on a run-time level as in our approach. However, despite the intention to apply generic frameworks for the specification, analysis, and verification of object-oriented concurrent programs, e.g.~in \cite{ZambonR11,DottiDFRRS05}, there are no existing publicly available tools implementing this long-term goal that are powerful enough for \scoop.

\paragraph{Outlook}
Our current approach allows for automatically verifying \scoop programs, with the help of a simple toolchain consisting of a compiler from \scoop to a \gts-based run-time model that then can be analysed and verified with \groove. A streamlined instance of our toolchain, including a publicly available version of the compiler, will be available soon at~\cite{Referee-Webpage}. As already mentioned, our run-time model can be seen as another operational semantics for \scoop programs: a more detailed formal comparison with competing formalisations, e.g.~\cite{Morandi-SNM13a}, is currently on the way based on a more stringent formalisation of the \cpm model and its extensions.

The given verification approach and modelling choices can also be applied to other concurrent asynchronous libraries and languages, e.g.~the alternative concurrent Eiffel model CAMEO~\cite{Brooke-Paige09a}, and the existing \gts formalisation of Grand Central Dispatch~\cite{Geeraerts-HR13a}. As a future step, we want to include verification approaches beyond the strategies of \groove, which will depend on novel abstraction techniques for \cpm (and its extensions), e.g.~in the spirit of pattern abstraction \cite{Rensink-Zambon12a}, or cluster abstraction~\cite{BackesR15}. As a lot of verification properties can be boiled down to MSO properties on the underlying \gts, we also plan to enrich the verification techniques for concurrent asynchronous object-oriented programs with ideas from program logics for \gts, e.g.~as detailed in~\cite{Habel-Pennemann09a,Poskitt-Plump14a}.
We also plan to publish the current toolchain in a more convenient front end by providing a bridge from existing \scoop IDEs to \groove.


\bigskip
{\small \noindent \textbf{Funding/Acknowledgements.} The underlying research was partially funded by the
European Research Council under FP7/2007-2013 / ERC Grant agreement no.\ 291389. We thank B.\ Meyer and the \scoop community  for valuable ``scooped'' discussions and A.\ Rensink for feedback on the internals of \groove. Finally, we thank the anonymous GaM referees for their insightful comments, which helped to improve this paper.

\bibliographystyle{eptcs}
\bibliography{references}

\end{document}